*Classification:* Physical Sciences, Applied Physical Sciences

Motionless Phase Stepping in X-Ray Phase Contrast Imaging with a Compact Source

Houxun Miao[a], Lei Chen[b], Eric E. Bennett[a], Nick M. Adamo[a], Andrew A. Gomella[a], Alexa M. DeLuca[a], Ajay Patel[a], Nicole Y. Morgan[c], and Han Wen[a, 1]

[a] Imaging Physics Laboratory, Biochemistry and Biophysics Center, National Heart, Lung and Blood Institute, National Institutes of Health, Bethesda, MD 20892

[b] Nanofab, Center for Nanoscale Science and Technology, National Institute of Standards and Technology, Gaithersburg, MD, 20899

[c] Microfabrication and Microfluidics Unit, Biomedical Engineering and Physical Science Shared Resource, National Institute of Biomedical Imaging and Bioengineering, National Institutes of Health, Bethesda, MD 20892

[1]To whom correspondence should be addressed. E-mail: wenh@nhlbi.nih.gov

**Abstract**

X-ray phase contrast imaging offers a way to visualize the internal structures of an object without the need to deposit any radiation, and thereby alleviate the main concern in x-ray diagnostic imaging procedures today. Grating-based differential phase contrast imaging techniques are compatible with compact x-ray sources, which is a key requirement for the majority of clinical x-ray modalities. However, these methods are substantially limited by the need for mechanical phase stepping. We describe an electromagnetic phase stepping method that eliminates mechanical motion, and thus removing the constraints in speed, accuracy and flexibility. The method is broadly applicable to both projection and tomography imaging modes. The transition from mechanical to electromagnetic scanning should greatly facilitate the translation of x-ray phase contrast techniques into mainstream applications.

Today x-ray imaging modalities accounts for the majority of the diagnostic imaging procedures in the United States (1). Conventional x-ray images depict variations in the attenuation of transmitted x-rays according to the density distribution in the body. The attenuation-based contrast is typically low in soft tissues at the photon energies commonly employed in medical imaging (25-100 keV). However, the phase shift for incident x-rays is many times the linear attenuation of the amplitude in weakly absorbing materials, which points to phase contrast imaging as a means for better resolving soft tissue structures without the need to deposit the harmful radiation (2). Among the first demonstrations of phase contrast in x-ray imaging were diffraction enhanced images using Bragg analyzer crystals (3, 4) and free space propagation of transversely coherent waves (5, 6). In the mid-1990s monolithic crystal interferometers (7) were used to obtain the first absolute phase shift images of soft tissue samples (2). A few years later, x-ray differential phase contrast (DPC) imaging with a grating interferometer was proposed and then realized (8-10). The grating Talbot interferometer works with a broader energy band and is less affected by environmental changes than monolithic crystal interferometers, at the cost of lower phase sensitivity (11). A major improvement in grating-based imaging was quantitative phase retrieval by the phase stepping method (10, 12-14). The Talbot-Lau interferometer enabled the use of commercial x-ray tubes for grating DPC, by adding a source grating to give spatial coherence to an extended source (8, 15). A preclinical x-ray DPC computed tomography (CT) scanner incorporating the above advances has been recently demonstrated (16).

Despite the rapid progress in grating-based phase contrast imaging, there are fundamental challenges when it comes to routine applications outside the laboratory environment. Its Achilles' heel is the mechanical phase stepping process, in which one grating is physically moved in multiple steps over a grating period in order to obtain a single differential phase image (10, 14). Accurate mechanical movement of centimeter-size objects at the sub-micron level is inherently slow and difficult to reproduce precisely without a static and stabilized platform. In common configurations including fluoroscopes and CT scanners, the precision motors must be mounted on moving gantries, which will lead to additional mechanical instability (16). Mechanical phase stepping may also be the ultimate limit of the imaging rate, which is critically important for fluoroscopy and clinical CT scans. Recognizing this, two methods for improving the imaging speed in phase contrast CT applications have been reported. One method combines a pair of projection images taken from opposing directions to retrieve phase information, but assumes weak phase objects, no scattering, and a constant, uniform background phase over the whole field of view (17). The second method reduces the number of phase steps by sharing them among neighboring projection angles in a CT scan, but still relies on mechanical phase stepping (18). A more general alternative to phase stepping is the Fourier fringe analysis method, which extracts phase information from a single image containing interference fringes, but with reduced spatial resolution (19-25).

A basic solution to the limitations of mechanical phase stepping should remove the need for physical movement completely. In the fields of radar and ultrasonic imaging, the development of electronic beam steering, which replaced mechanical scanning of the antenna or probe, greatly improved the speed and capability of both technologies (26, 27). We report on an analogous solution for grating-based x-ray phase contrast imaging called electromagnetic phase stepping (EPS). We created an adaptive processing algorithm as part of the method, and demonstrate its effectiveness in imaging studies of rodents and other samples in a benchtop system.

**Results**

A generic grating-based phase contrast imaging system consists of an x-ray tube, a Talbot-Lau interferometer and an x-ray camera as schematically illustrated in Fig. 1A. The interferometer has two amplitude gratings ($G_0$, $G_2$) and one phase grating ($G_1$). In our system the grating period is 4.8 μm. Grating $G_0$ splits the x-ray cone beam into a number of thin line sources whose lateral coherent lengths are greater than the grating period at the plane of $G_1$. Each line source creates an intensity fringe pattern, i.e. fractional Talbot image (10, 28) of $G_1$, on the plane of $G_2$. Because the fringe period is usually smaller than the detector resolution, $G_2$ is used to produce a broader moiré pattern. When the distance between $G_0$ and $G_1$ is the same as that between $G_1$ and $G_2$, the fringe pattern from each individual line source adds up constructively on the plane of $G_2$.

If $G_0$ and $G_1$ are parallel and $G_2$ is rotated around the optical axis with respect to $G_1$ by a small angle $\theta$, the differential phase information is encoded into the moiré fringes on the detector plane:

$$I \approx a_0 + a_1 \cos\left[\frac{2\pi}{p}\left(x\theta + \frac{\lambda d}{\pi}\frac{\partial \Phi}{\partial y}\right) + \phi_b\right], \tag{1}$$

where $x$ and $y$ are coordinates in the detector plane, $a_0$ is the un-modulated baseline, $a_1$ is the fringe amplitude, $p$ is the grating period, $d$ is the distance between $G_1$ and $G_2$, $\lambda$ is the x-ray wavelength, and $\phi_b$ is the background instrumental phase, which depends on the positions of the gratings. The desired information is the derivative of the x-ray phase shift caused by the sample, expressed as $\partial \phi/\partial y$ in the detector plane.

The phase stepping method calculates the differential phase image from several images with different background phases $\phi_b$. To date this has required physically moving one of the gratings in the $y$ direction over multiple steps that cover a grating period, while taking an image at each step. In this process the moiré pattern in the images visibly moves across the static projection of the object, giving rise to the intuitive term of "fringe scanning" as a synonym of phase stepping.

Recognizing that the essential requirement of fringe scanning is a relative movement between the moiré fringes and the projection image of the object, electromagnetic phase stepping achieves the condition by electromagnetically shifting the focal spot of the x-ray tube in a transverse direction across the fringe pattern, e.g. with an externally applied magnetic field that deflects the electron beam in the x-ray tube (Fig. 1A). Shifting the focal spot causes an opposite movement of the projection of the object on the detector plane, while the fringes can be made to remain stationary or move by a different amount. In our setup of the Talbot-Lau interferometer, the moiré fringes are produced by a slight rotation of the third (analyzer) grating. In this case the fringes remain stationary in spite of the shifting focal spot. In the inverted embodiment where the moiré fringes are produced by rotating the first (source) grating, the movement of the fringes will exceed that of the projection image. In all cases, the images can then be digitally shifted back to re-align the projections of the object. The result is that the fringes move over a stationary projection image, effectively synthesizing the phase stepping process (Fig. 1B).

It is worth noting that shifting the focal spot of the cone beam also causes a slight change of the projection angle on the object. This change is negligible for objects that occupy a small fraction of the distance between the source and the camera. When the object thickness is a significant fraction of the source-camera distance, the reconstruction algorithm takes on the characteristics of stereoscopic imaging or tomosythesis. A more detailed discussion is provided in the Supporting Information.

In our imaging device the gratings are rigidly mounted. A solenoid coil is attached to the front surface of the x-ray source (Fig. 1C) and is driven by a 25 volt power supply to produce the magnetic field with full digital control and a response time of 200 microseconds. The resulting focal spot shift causes an opposite movement of the object projection on the camera over a stationary fringe pattern (see Movie S1 in Supporting Information). A magnetic field of 2.4 mT was sufficient to shift the projected image over one period of the moiré fringes (300 μm). The details of the experimental setup are described in the Methods.

We created an adaptive image processing algorithm to extract the differential phase contrast, scatter (dark-field) (20, 29) and conventional attenuation images from the EPS data without prior knowledge of the movement of the object projection or of the moiré fringes (see Supporting Information). The algorithm first aligns the projections of the object, generating an image stack in which the object is stationary while the moiré fringe pattern moves over it. The aligned images are then processed as fringe-scanned images by the second part of the algorithm, which measures the phase increments in the phase stepping process with a Fourier transform method(2, 19, 20, 30). The algorithm can cope with arbitrary and spatially varying phase increments without assuming a priori knowledge, and thus is robust against potential instabilities in the alignment of various components in the imaging system.

Using the benchtop system with electromagnetic phase stepping, we first imaged a reference sample containing borosilicate spheres of 5 mm diameter. The full series of images is included in Movie S1 in the Supporting Information. Figure 2 shows the processed DPC, absolute phase shift (via direct integration of the DPC), and linear intensity attenuation images. The stripe artifacts in the phase shift image result from the lack of low spatial frequency information in the DPC signal. The linear attenuation at the sphere centers is 0.61±0.02, corresponding to an effective mean x-ray energy of 35.5±0.5 keV. The total phase shift at the sphere centers is estimated to be $(3.2±0.2) \times 10^2$ rad, from which the real part of the refractive index decrement of the borosilicate material is estimated to be $\delta = (3.6±0.3) \times 10^{-7}$, in good agreement with the reference value of $3.7 \times 10^{-7}$ for 35.5 keV x-rays (The refractive index is calculated from the X-ray form factor, attenuation and scattering tables available at http://physics.nist.gov/PhysRefData/FFast/html/form.html).

As an example of a biological specimen, Fig. 3 shows the DPC and linear attenuation images of a cricket obtained by electromagnetic phase stepping. The DPC image reveals more detailed structures throughout the head, the body and the legs of the cricket, owing to its sensitive nature to small changes in the density of the constituents.

Biomedical research routinely involves the imaging of rodents, and the photon energy of the benchtop system was sufficient to penetrate the body of adult mice. As examples, we imaged the head region of a euthanized mouse in air and the torso region of another euthanized mouse fixed in 10% buffered formalin.

Figures 4A-D are the processed images of the head region, including the DPC, phase-contrast enhanced, linear intensity attenuation, and scatter (dark-field) linear extinction images. The DPC signal degrades into random phase noise in the areas of the metallic ear tag due to the strong attenuation of the fringes. Phase retrieval by direct integration of the DPC image can lead to substantial errors for reasons. The first relates to the inherent lack of low frequency information in the DPC data, and the second factor is the random phase values in areas where the interference fringes are strongly suppressed, either due to attenuation (as around the metallic ear tag) or scattering. The first can be addressed by the method of Roessl and co-authors (31), which merges the low spatial frequency content of the intensity attenuation with the high spatial frequency content of the DPC data according to the scaling between the real and imaginary parts of the refractive index of the material (here, soft tissue). The second problem is circumvented by substituting the derivative of the intensity attenuation into the DPC image in areas where the DPC information is missing, again using an appropriate scaling factor. These calculations are described in some detail in Methods. The end result is a phase-contrast enhanced image (Fig. 4B). As indicated by the white arrows, numerous details emerge in the phase-contrast enhanced image which cannot be clearly seen in the intensity attenuation image.

Figure 5 is a compilation of reconstructed images of the torso region of the mouse. The DPC image highlights weakly absorbing phase objects such as air bubbles (Fig. 5A). The phase-contrast enhanced image (Fig. 5B) contains both the phase and attenuation information. The lungs are most visible in the dark-field (scatter) image (Fig. 5C), owing to their porous microstructures. The high-density bones and the metallic ear tag are clearly visible in the intensity attenuation image (Fig. 5D).

Movies of the mouse head and torso images are provided in the Supporting Materials for comparison between phase-contrast enhanced and linear intensity attenuation images.

**Discussion**

Grating interferometers used with phase stepping enable high-resolution x-ray phase contrast imaging with compact x-ray tubes. However, the stringent requirements for mechanical phase stepping have been a major challenge in bringing phase contrast into common imaging systems. The electromagnetic phase stepping method and the adaptive processing algorithms presented here effectively replace the precision mechanical scanning system and its associated engineering challenges with a simple solenoid coil attached to the x-ray source, providing substantial advantages in speed, accuracy and flexibility. The near instantaneous control of the focal spot could also enable real-time compensation for instrumental instabilities, including thermal drift and vibrations. The transition from mechanical to electromagnetic scanning also reduces the cost of parts and maintenance and should improve reliability, all of which may contribute to the translation of phase contrast techniques into mainstream applications. For biomedical imaging, grating periods of a few hundred nanometers are being developed for greater phase sensitivity (32, 33). Here, electromagnetic phase stepping may become a necessity, as precise mechanical movement at the nanometric level may be difficult to achieve outside the most favorable settings.

**Methods**

**Technical Specifications of the Imaging System**

We used a tungsten-target x-ray tube (SB-80-1k, Source Ray Inc.) operating at a peak voltage of 55 kV and a current of 1mA as the source. The focal spot of the tube was approximately 50 μm. The Talbot-Lau interferometer consisted of three gratings of 4.8 μm period (Fig. 1A). Gratings $G_0$ and $G_2$ were intensity modulating (amplitude) gratings, $G_1$ was a phase grating. The grating lines were oriented horizontally. The amplitude gratings (Microworks GmbH) had gold-filled trenches of 60 μm nominal depth in a polymer substrate (34, 35). They were rotated around the vertical axis by 45° to increase the effective gold height(36). The phase grating had un-filled trenches etched into a silicon substrate using the Bosch process (37), with an etch depth of 27 μm. It was also rotated by 45° to be parallel with the other gratings. The gratings were positioned at equal spacing over a total distance of 76 cm. The third grating ($G_2$) was slight rotated around the optical axis to create vertical moiré intensity fringes of approximately 300 μm period on the detector plane. With this arrangement the moiré fringes are largely independent of the position of the x-ray source (see Movie S1 for a demonstration). The x-ray camera (PI-SCX-4096, Princeton Instruments) had a pixel size of 30 μm and a pixel matrix of 2048x2048.

For electromagnetic phase stepping, a home-made solenoid coil of copper wire (60 mm inner diameter, 200 turns) was attached to the front surface of the x-ray tube housing (Fig.1B). The coil was driven by a digital power supply which provided up to 2.0 A of current at up to 8 W of power. The corresponding peak magnetic field was calculated to be 3.1 mT at the location of the electron beam inside the x-ray tube. The field strength was verified experimentally with a magnetometer. The electron beam is oriented vertically. The magnetic field shifted the focal spot by up to 380 μm (with 1.5 A current applied) in the horizontal direction, perpendicular to the moiré fringes. The deflections of the focal spot at various levels of input current into the coil were measured experimentally as shown in Fig. S1 in the Supporting Information.

**Image Acquisition and Processing in Electromagnetic Phase Stepping**

The method of electromagnetic phase stepping obtains three types of information from a single set of raw images: the differential phase, the conventional linear attenuation, and the dark-field (scatter) images. For the method to be broadly applicable, the image reconstruction algorithm needs to be able to cope with a number of instabilities that may be present in a compact imaging device. These include drift of the focal spot of the x-ray tube, drift in the alignment of the gratings and other components, and variable positioning of the imaged object.

Each phase stepping cycle includes 6 progressive levels of input current into the field coil from 0 to 1.5 A, resulting in six different positions of the cone beam focal spot (Fig. 1A). At each position an image is taken. The image contains a moiré fringe pattern, which is modulated in amplitude and in phase by the projection of the object. A shift of the focal spot results in a displacement of the projection in the opposite direction, as well as a change of the projection angle. The moiré fringes remain stationary with the specific arrangement of the gratings in our device described above. If we make the assumption that the thickness of the object takes up a small fraction of the distance between the source and the camera, the movement of the projection is approximately uniform throughout the sample, and the change in projection angle can be neglected. More generally, the movement of the projection of a given transverse plane across the beam axis (focal plane) is determined by its position between the source and the detector. Therefore, the image reconstruction effectively focuses on a transverse slice through the object. This is similar to tomosynthesis.

The reconstruction algorithm is adaptive in two ways: first, owing to variable sample positioning, the movement of the projection of the object is not assumed in advance but determined retrospectively from the images themselves; and second, once the projections are aligned, the algorithm needs to retrieve phase and amplitude images from a set of arbitrary fringe positions, i.e. non-uniform and spatially varying phase increments between successive images in a phase stepping set.

The displacement of the sample projection on the detector plane is measured through a Fourier space analysis that demodulates the moiré fringes while retaining the projection image at a reduced resolution (2, 19, 20). The relative movement of the projections are measured from the demodulated images. The full-resolution images are then shifted by the opposite amounts to align all the projections. In the aligned images, the projection of the object remains static while the moiré fringes move across it. The intensity at each pixel oscillates with the moving fringes. In effect, the aligned images are equivalent to images acquired in mechanical phase stepping by moving one of the gratings. The details of the reconstruction procedure after the alignment step are described in the Supporting Information.

**Reconstruction of Phase Contrast Enhanced Images**

The algorithm is as follows: if we define $A_0$ and $\Phi$ as the linear attenuation and the phase shift of the x-ray wave front after propagation through the object, $A_0$ is simply the absolute value of the natural logarithm of the transmission, and $\frac{\partial \Phi}{\partial y}$ corresponds to the DPC signal. The first step is to incorporate the derivative of the linear attenuation $\frac{\partial A_0}{\partial y}$ into the DPC signal in a weighted sum of $\frac{\partial A_c}{\partial y} = CW_0 \frac{\partial A_0}{\partial y} + W_1 \frac{\partial \Phi}{\partial y}$, where $C$ is the scaling factor between the real and imaginary parts of the refractive index (31), and the weights $W_0$ and $W_1$ are determined locally according to the amplitude of the interference fringes $A_1$ and the noise level $N_1$ in the fringe amplitudes. Specifically, $W_0(r) = 1/\{1 + [A_1(r)/(2N_1)]^6\}$, and $W_1(r) = 1-W_0(r)$, Once the combined differential image $\frac{\partial A_c}{\partial y}$ is determined, it is merged with the intensity attenuation data into a phase contrast enhanced image according to the algorithm described in Ref 31: the direct integral of $\frac{\partial A_c}{\partial y}$ is high-pass filtered in the Fourier space, and merged with the low spatial frequency part of the linear intensity attenuation, then inverse Fourier transformed into the final image.

**Acknowledgements**

We are grateful to the staff of the Nanofab facility of National Institute of Standards and Technology, Gaithersburg, Maryland, for assistance with fabrication of the phase grating, and to Prof. Marco Stampanoni of Paul Scherrer Institute for helpful discussion on the fusion of phase and intensity images. This work was supported by the Intramural Research Program of the National Institutes of Health, including the National Heart, Lung, and Blood Institute and the National Institute of Biomedical Imaging and Bioengineering.

**Figure Legends**

Fig. 1. Schematic illustration of the electromagnetic phase stepping apparatus and principles. (*A*) The imaging system (not to scale) is a standard grating interferometer, but with a solenoid coil attached to the x-ray tube housing. The energized coil produces a magnetic field and an associated Lorenz force on the electron beam in the x-ray tube, shifting the impact spot on the anode target from which x-rays are emitted. Correspondingly, the projection image of the sample is displaced in the opposite direction on the camera (displacement exaggerated for better viewing). (*B*) Schematic representation of electromagnetic phase stepping. The top red curve represents the moiré fringe pattern, and the bottom red curve represents the projection of the object. The red triangle and circle represent a specific location in the projection and its position in the fringe pattern, respectively. In conventional phase stepping (PS), moving one grating shifts the fringe pattern to a new position represented by the dashed blue line. Consequently, the same location in the object projection has a different phase, represented by the blue circle on the dashed blue line. In electromagnetic phase stepping (EPS), the object projection is shifted (dashed green line) due to the displacement of the focal spot of the cone beam, moving the same location in the projection from the red triangle to the green triangle, with a corresponding shift on the fringe pattern (red curve) from the red circle to the green circle. Effectively, the same physical location in the object projection is given a phase shift. (*C*) Photographs of the home-made solenoid coil and its placement around the cone-beam window of the x-ray tube housing.

Fig. 2. Images of a reference sample containing borosilicate spheres. (*A*) Differential phase contrast, (*B*) phase shift, obtained by direct integration of the DPC information and baseline corrected through linear fitting, (*C*) linear intensity attenuation and (*D*) a cross section profile of the phase shift through the center of a sphere, the location of which is marked by the green line in (*B*). The blue and red curves are experimental and modeled data for a sphere, respectively.

Fig. 3. Reconstructed images of a cricket. (*A*) Linear intensity attenuation, (*B*) differential phase contrast. A tungsten bead of 0.8 mm diameter (bright dot in *A*) was placed near the head of the cricket as a marker.

Such markers are used to accurately determine the displacement of the projection images during electromagnetic phase stepping.

Fig. 4. Reconstructed images of the head region of a mouse. (*A*) Differential phase contrast, (*B*) phase-contrast enhanced, (*C*) dark-field (scatter), and (*D*) linear intensity attenuation. Arrows in (*B*) indicate examples of features more visible in the phase-contrast enhanced image than in the classic intensity attenuation image (*D*). The bright U-shaped object is a metallic ear tag.

Fig. 5. Reconstructed images of the torso region of a mouse. (*A*) Differential phase contrast, (*B*) phase-contrast enhanced, (*C*) dark-field (scatter), and (*D*) linear intensity attenuation. The lungs are most clearly seen in the scattering image (*C*).

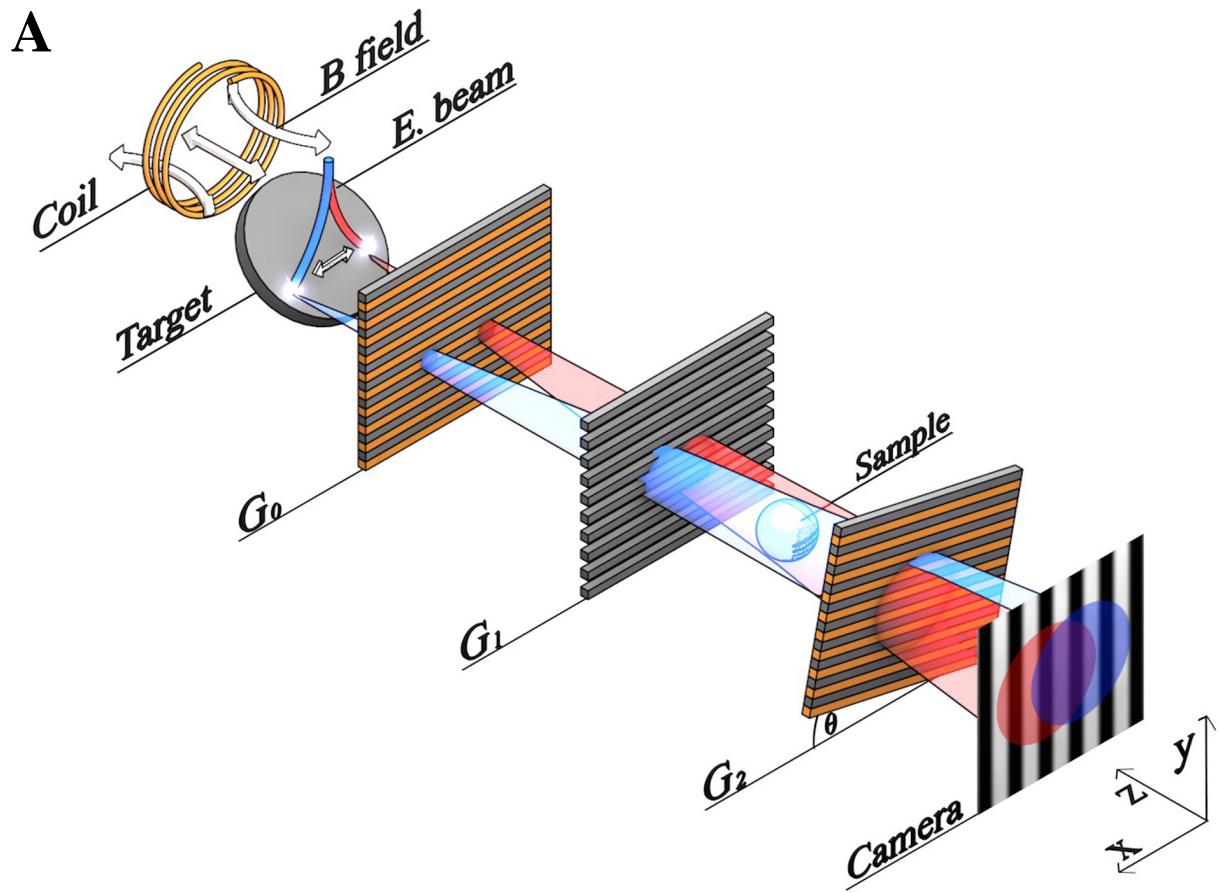

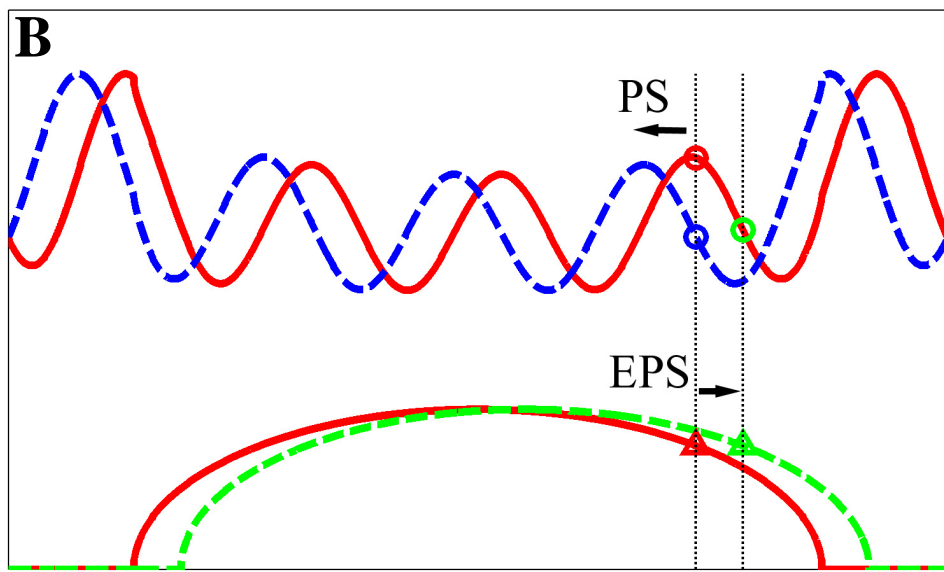
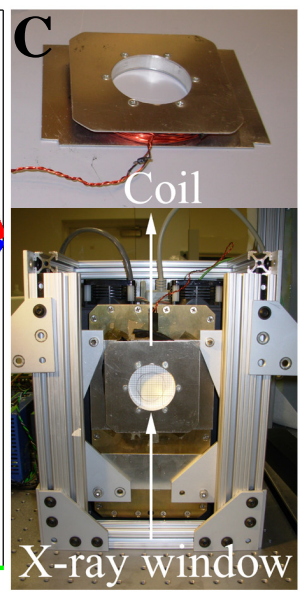

Fig. 1

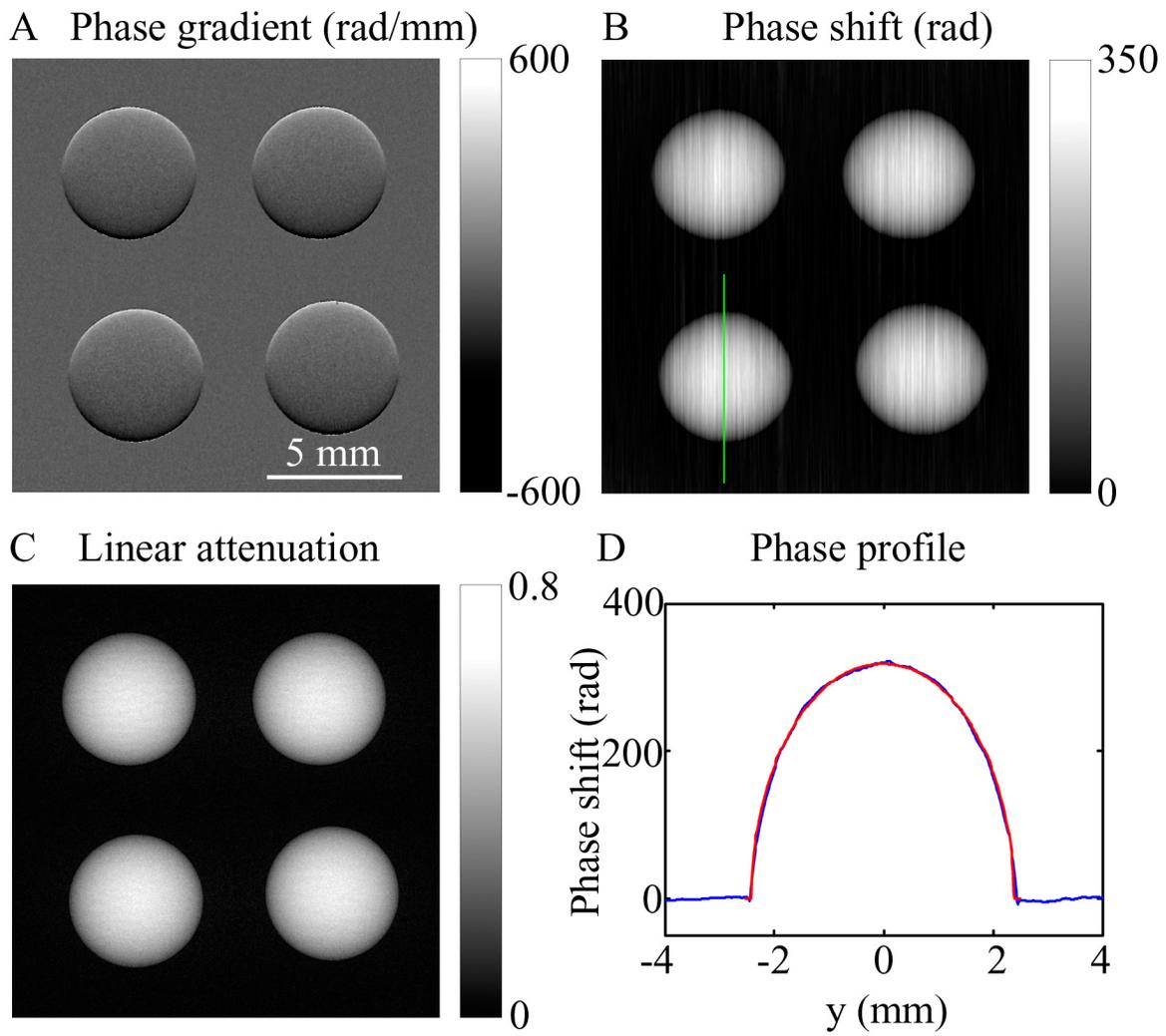

Fig. 2

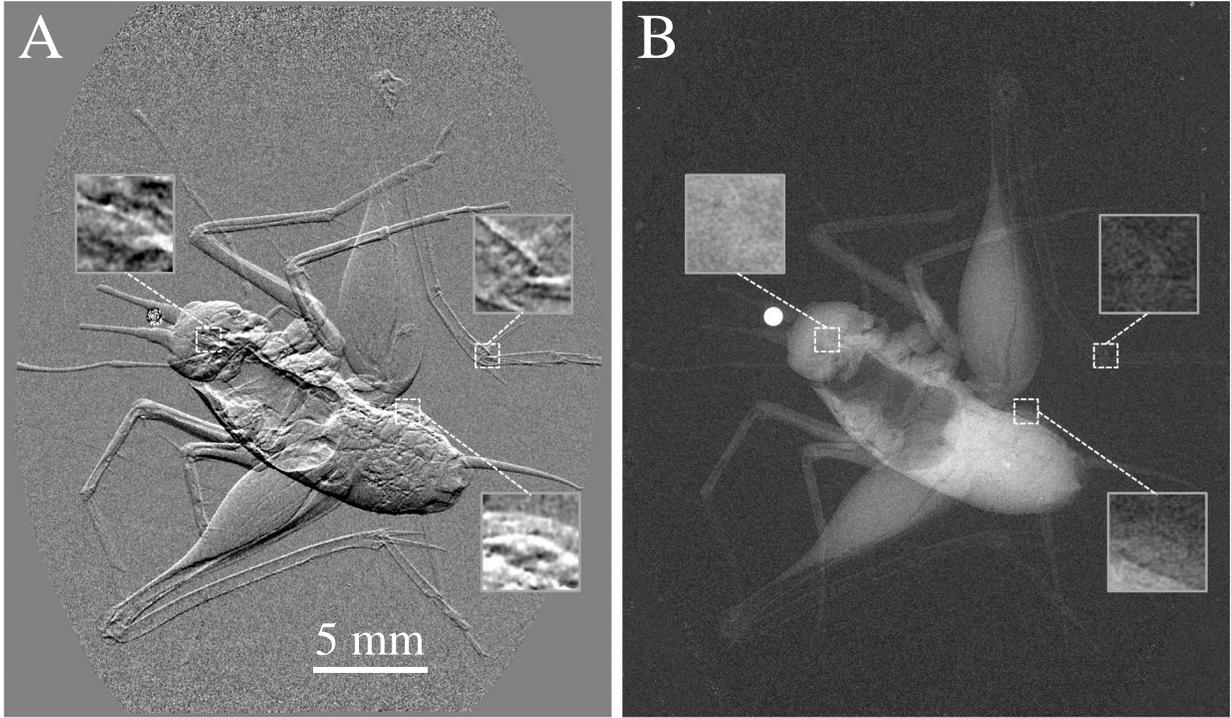

Fig. 3

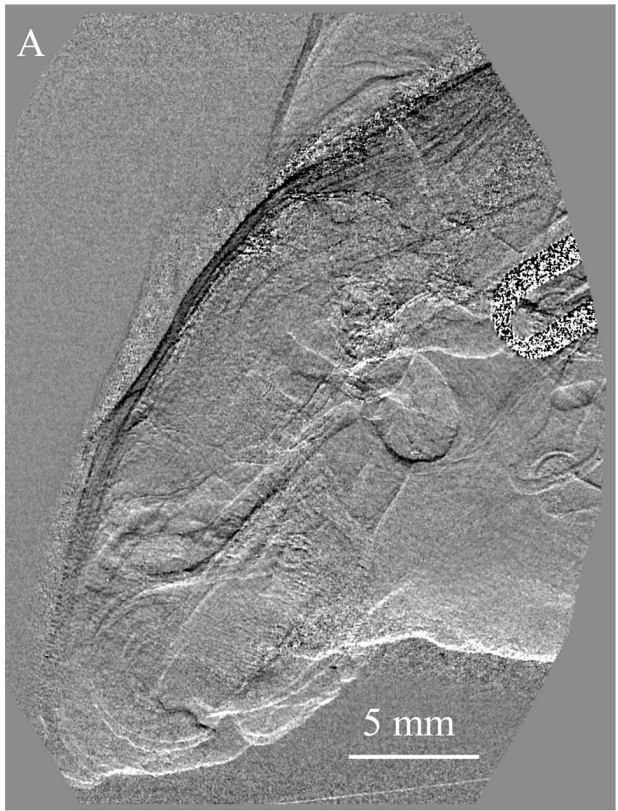
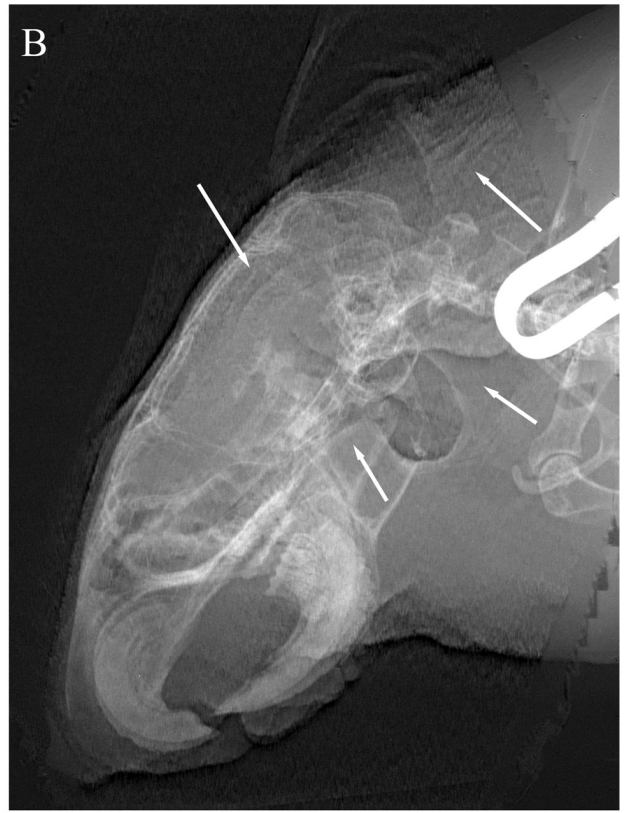
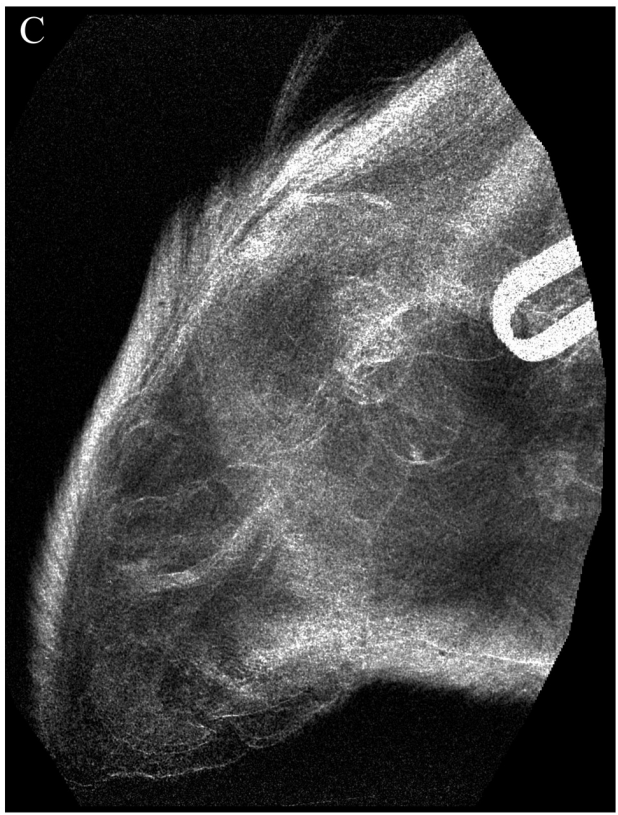
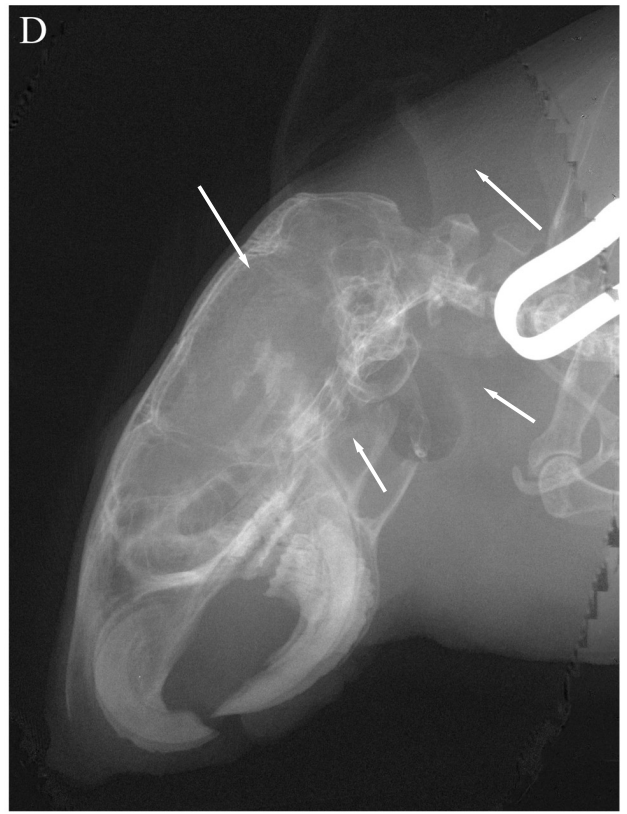

Fig. 4

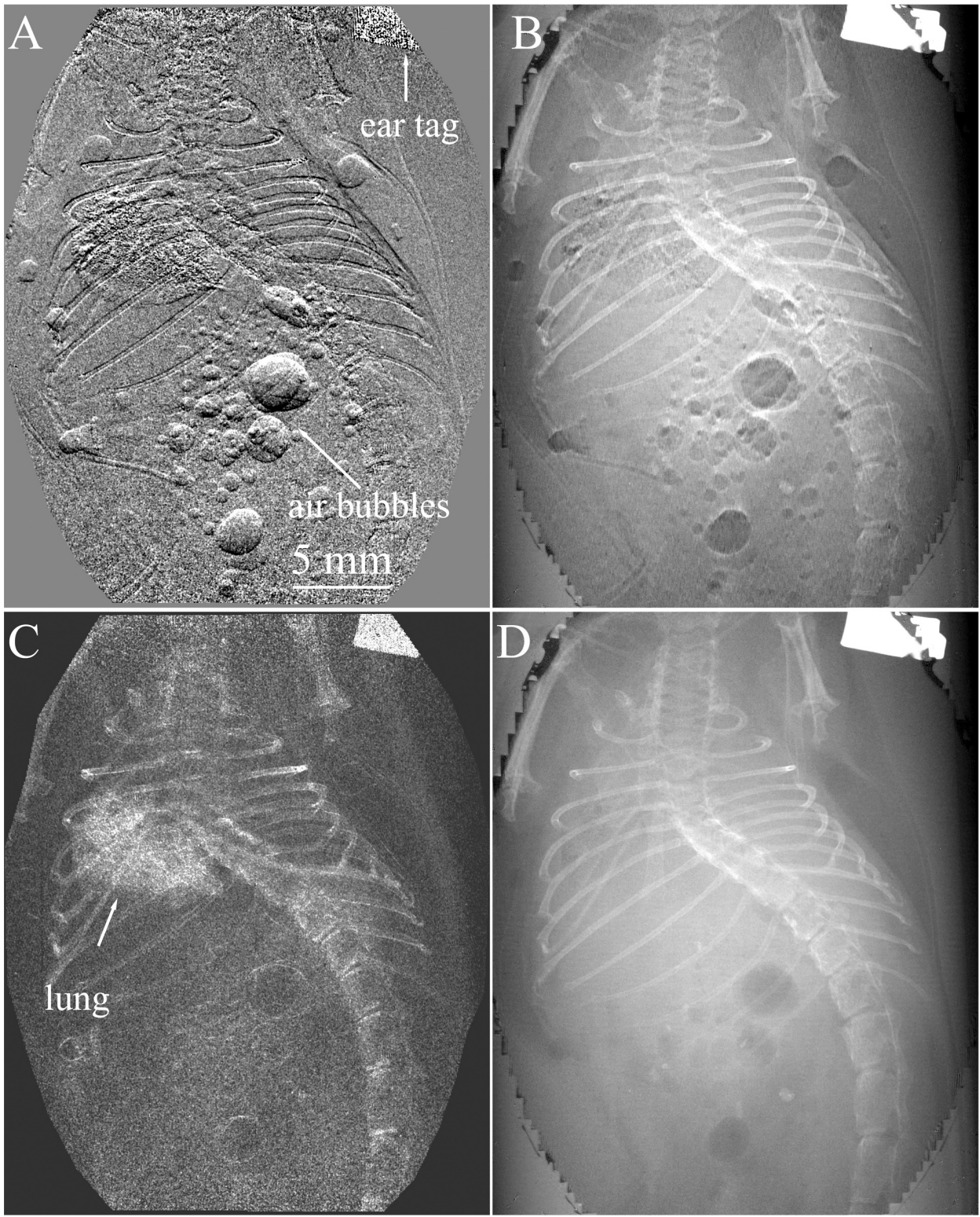

Fig. 5

Motionless Phase Stepping in X-Ray Phase Contrast Imaging with a Compact Source

Houxun Miao[a], Lei Chen[b], Eric E. Bennett[a], Nick M. Adamo[a], Andrew A. Gomella[a], Alexa M. DeLuca[a], Ajay Patel[a], Nicole Y. Morgan[c], and Han Wen[a, 1]

[a] Imaging Physics Laboratory, Biochemistry and Biophysics Center, National Heart, Lung and Blood Institute, National Institutes of Health, Bethesda, MD 20892

[b] Nanofab, Center for Nanoscale Science and Technology, National Institute of Standards and Technology, Gaithersburg, MD, 20899

[c] Microfabrication and Microfluidics Unit, Biomedical Engineering and Physical Science Shared Resource, National Institute of Biomedical Imaging and Bioengineering, National Institutes of Health, Bethesda, MD 20892

[1] To whom correspondence should be addressed. E-mail: wenh@nhlbi.nih.gov

# Supporting Information

**Image Retrieval Algorithm**

The algorithm is adaptive in two aspects: owing to variable sample positioning, the movement of the projection of the object in the electromagnetic phase stepping (EPS) process is not known *a priori*, and needs to be determined retrospectively from the images themselves; once the projections are aligned, the algorithm needs to retrieve phase and amplitude images from a set of arbitrary and spatially varying phase increments between successive images of a phase stepping set. The following is an outline of the algorithm.

Noting the spatial frequency of the moiré fringes as $g$, the aligned $k$th image of the EPS set can be expressed generally as

$$I_k(r) = A_0(r) + \sum_{n>0} A_n(r) \cos\{n[2\pi g x + D(r)] + b_{n,k}(r)\},$$

where $r$ represents two-dimensional coordinates in the image plane, $x$ is the coordinate perpendicular to the moiré fringes, $n$ labels the harmonic component centered around the spatial frequency that is $n$ times the fundamental frequency of the moiré fringes, $A_0(r)$ is the magnitude of the smoothly varying part of the image intensity, $A_n(r)$ are the amplitudes of the rapidly varying components associated with the moiré fringes, $D(r)$ is the differential phase signal from the object, and lastly, $b_{n,k}(r)$ represents the phase shift of the moiré pattern in the $k$th image plus all other instrument related phase contributions. The phase $b_{n,k}(r)$ is the only quantity that is dependent on the phase step counter $k$.

For clarity, the discussion below will focus on the common situation where only the first harmonic of $n = 1$ has a significant magnitude. The task of image processing is to obtain $A_0(r)$, $A_1(r)$ and $D(r)$. These can be calculated from the set of EPS images, if the instrument phase $b_{1,k}(r)$ is first extracted.

The instrumental phase $b_{1,k}(r)$ is obtained through a Fourier transform method (1–4) that demodulates the fringes by way of windowed filters in the Fourier space. The result is a phase image $D(r) + b_{1,k}(r)$ at a reduced resolution which is approximately equal to the period of the moiré fringes. The difference of the results between the $k$th and the 0th images is $b_{1,k}(r) - b_{1,0}(r)$, again at the reduced resolution. If the periods of the gratings are uniform in space or vary gently, then the phase difference of $b_{1,k}(r) - b_{1,0}(r)$ also varies smoothly in space. We can then interpolate the low-resolution version in space to obtain $b_{1,k}(r) - b_{1,0}(r)$ at the full resolution of the detector pixel size. At this point, we have sufficient information to extract $A_0(r)$, $A_1(r)$ and the phase $D(r) + b_{1,0}(r)$ on a pixel-by-pixel basis by a least-squares fitting procedure.

At this stage there remain two issues to resolve. The first is to remove the instrumental contribution of $b_{1,0}(r)$ from the resulting phase map. This is complicated by the second problem, namely the difficulty in obtaining the instrument phase for parts of the object where the x-rays are strongly scattered or attenuated, causing the moiré fringes to vanish.

Both problems are solved by acquiring a reference EPS set without any sample. By the same Fourier analysis the instrumental phase differences $b'_{1,k}(r) - b'_{1,0}(r)$ are extracted, and the amplitudes of $A'_0(r)$, $A'_1(r)$ and the instrumental phase $b'_{1,0}(r)$ are readily obtained (since $D(r)$ is zero). The problem of vanishing fringes in some areas of the sample is solved by substituting the reference instrument phase $b'_{1,k}(r) - b'_{1,0}(r)$ for its counterpart $b_{1,k}(r) - b_{1,0}(r)$ with the sample present, plus a correction that accounts for any instrumental drifts that may occur between the acquisitions of the reference and sample images. The correction is obtained by a linear fitting of the difference over areas where the moiré fringes are well defined. The substitutions allow robust calculations of $A_0(r)$, $A_1(r)$ and $D(r) + b_{1,0}(r)$ across the full field of view.

The last step is to remove the instrument phase $b_{1,0}(r)$ from the above result. Here again the reference instrument phase $b'_{1,0}(r)$ is used to substitute $b_{1,0}(r)$.

The absorption image of the object is simply the ratio of $A_0(r)/A'_0(r)$. The scatter or dark-field image measures additional attenuation of the interference fringes due to scattering in the object. It is given by $[A_1(r)/A'_1(r)]/[A_0(r)/A'_0(r)]$.

So far it is assumed that the thickness of the object is small relative to the distance between the source and the detector. Under this assumption, the small change of the projection angle associated with a shift of the source point has negligible effect. In the opposite situation where the object occupies a large portion of the distance between the source and the detector, the change of projection angle needs to be taken into account. The way to do so depends on the mode of imaging. In planar imaging, scanning the source point provides data similar to those of tomosynthesis. The above reconstruction algorithm for EPS is based on translating the raw images by specific distances to align the object projection on the detector plane. This is valid for a transverse plane (focal plane) at a specific distance from the source point. The result will have a depth-of-focus character, where the slice at the focal plane has the sharpest resolution. Features become blurred in front of and behind the focal plane. In three-dimensional imaging by computed tomography, the imaging system is rotated around the object to cover a range of projection angles. In this case, the actual angle of the projection images should take into account the source point shifts. The corrected projection angles can then be used in the subsequent reconstruction steps.

**Deflections of the X-Ray Focal Spot**

Figure S1 shows the deflections of the x-ray focal spot as a function of the input current into the solenoid coil. The squares are experimentally measured values and the solid line is a linear fit of the measurement.

**Supporting Movies**

Movie S1 consists of a set of electromagnetically phase stepped images of borosilicate spheres. It shows the movement of the projections of the spheres over a static background of the moiré fringes, as the focal spot of the x-ray tube is shifted by the applied magnetic field. Movie S2 and S3 contain the intensity linear attenuation and the phase-contrast enhanced images of the two mice, for the purpose of comparison.

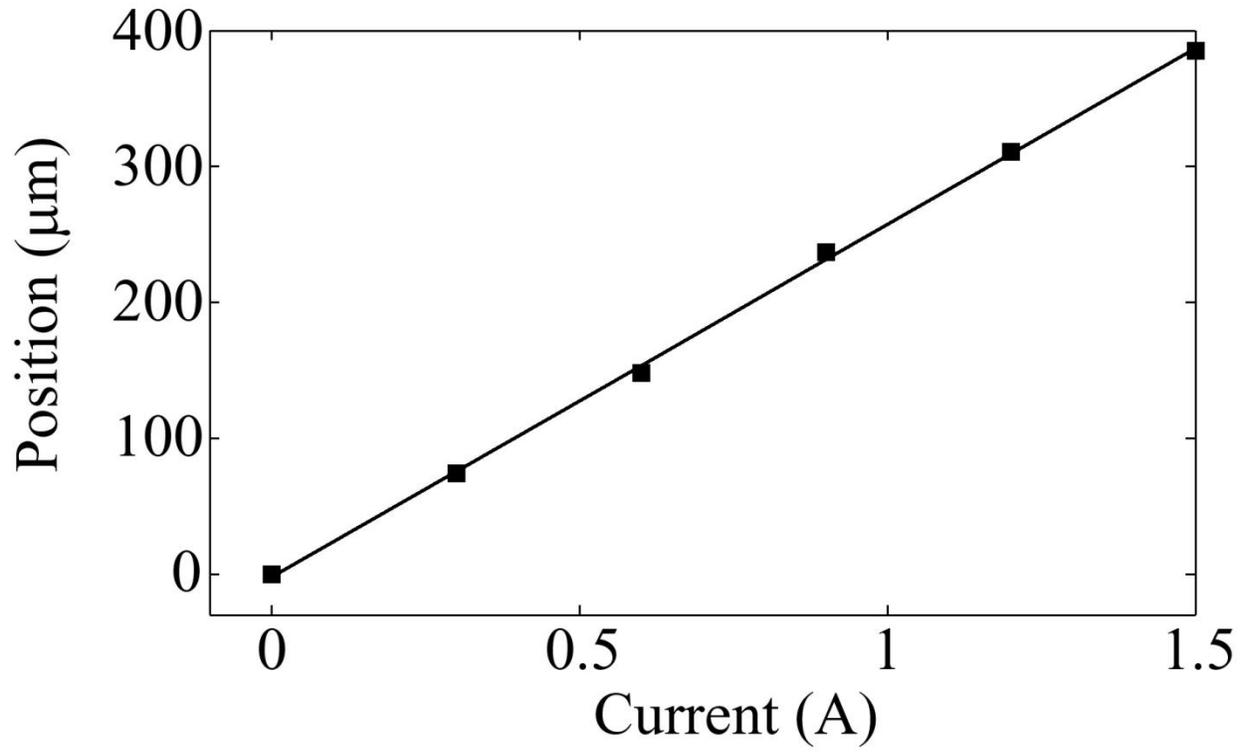

Fig. S1. Deflection of x-ray focal spot vs applied current.